\documentclass[a4paper,aps,twocolumn,showpacs,showkeys,nofootinbib,preprintnumbers,superscriptaddress,amsmath,amssymb,amsfonts]{revtex4-1}
\usepackage{graphicx}
\usepackage{bm,natbib}
\usepackage{amsmath}
\usepackage[latin1]{inputenc}
\usepackage[english]{babel}
\usepackage[T1]{fontenc}
\usepackage{amssymb}
\usepackage{amsfonts}
\usepackage{epsfig}
\usepackage{colordvi}
\usepackage{psfrag}
\usepackage{color}
\usepackage{dcolumn}
\usepackage{multirow}
\usepackage{hyperref}
\usepackage{epstopdf}
\usepackage{bm}
\usepackage{times}
\hypersetup{
  colorlinks=true,        
  linkcolor=blue,         
  citecolor=magenta,      
}

\newcommand{\aea}{Astron. Astrophys.}
\newcommand{\apjl}{Astrophys. J. Lett.}
\newcommand{\cqg}{Class.  Quant. Grav.}

\newcommand{\jcap}{J. Cosmol. Astropart. Phys.}
\newcommand{\mnras}{Mon. Not. R. Astron. Soc.}
\newcommand{\plb}{Phys. Lett. B}
\begin{document}

\title{Analysis of the Yukawa gravitational potential in $f(R)$ gravity II: \\ 
relativistic periastron advance}

\author{Mariafelicia De Laurentis}
\email{laurentis@th.physik.uni-frankfurt.de}
\affiliation{Institut f\"ur Theoretische Physik, Max-von-Laue-Stra{\ss}e 1, 
D-60438 Frankfurt am Main, Germany}
\affiliation{Dipartimento di Fisica "E. Pancini", Universit\'a di Napoli "Federico II", Compl. Univ. di Monte S. Angelo, Edificio G, Via Cinthia, I-80126, Napoli, Italy}
\affiliation{Lab.Theor.Cosmology,Tomsk State University of Control Systems and Radioelectronics(TUSUR), 634050 Tomsk, Russia}

\author{Ivan De Martino}
\email{ivan.demartino@ehu.eus}
\affiliation{Department of Theoretical Physics and History of Science,  University of the Basque Country UPV/EHU, 
Faculty of Science and Technology, Barrio Sarriena s/n, 48940 Leioa, Spain}

\author{Ruth Lazkoz}
\email{ruth.lazkoz@ehu.eus}
\affiliation{Department of Theoretical Physics and History of Science,  University of the Basque Country UPV/EHU, 
Faculty of Science and Technology, Barrio Sarriena s/n, 48940 Leioa, Spain}

\date{\today}

\begin{abstract}
Alternative theories of gravity may serve to overcame several shortcomings of the standard cosmological model but, 
in their weak field limit, General Relativity  must be recovered so as to match the tight constraints at the Solar System scale. 
Therefore, testing such alternative models at scales of stellar systems could give a unique opportunity to confirm or rule them out.
One of the most straightforward modifications is represented by analytical $f(R)$-gravity models
that introduce a Yukawa-like modification to the Newtonian potential thus modifying the dynamics of particles. 
Using the geodesics equations, we have illustrated the amplitude of these modifications. 
First, we have  integrated numerically the equations of motion showing the orbital precession of a particle around a massive object.
Second, we have computed an analytic expression for the periastron advance of systems having their semi-major axis much shorter than the Yukawa-scale length.
Finally, we have extended our results to the case of a binary system composed of two massive objects. 
Our analysis provides a powerful tool 
to obtain constraints on the underlying theory of gravity using current and forthcoming datasets.
\end{abstract}
\maketitle
\section{Introduction}
\label{uno}

Does General Relativity (GR) need to be modified to overcome the shortcomings at ultraviolet 
and infrared scales? This is one of the fundamental questions that still needs to be answered. 
As it is well known, GR is very well established  on the Solar System scale \cite{Will93, Stairs2003, Everitt2011},
and  forms the basis of the {\em concordance} cosmological model. Although, in the last decades 
many observational datasets have emerged confirming the model further \cite{Planck16_13,Perlmutter1997,Riess2004,Astier2006,Suzuki2012,Pope2004,Percival2001,Tegmark2004,Hinshaw2013,Blake2011,demartino2015b,demartino2016b}, 
some shortcomings have brought questions about whether GR is the true effective theory of gravity. 
First, GR is not a Quantum Theory and it cannot provide a description of the Universe
at quantum scales \cite{Misner1970, Isham1981}. Second,  GR cannot explain the emergence of the Large 
Scale Structure and the accelerated expansion of the Universe without adding two extra components to the total 
energy density budget, namely Dark Matter (DM) and Dark Energy (DE). 
The dynamical effects of these two components are evident at both galactic/extragalactic and cosmological scales, but  their fundamental nature, whether particles or scalar fields, is completely unknown  
\cite{Feng2010,Bertone2005,Capolupo2010,Schive2014,demartino2017b,Capolupo2017,demartino2018,Lopes2018,Panotopoulos2018}.
These problems have been interpreted as a breakdown of GR, and many alternative theories of gravity have been proposed \cite{darkmetric,Nojiri2011, PhysRept, idm2015, Nojiri2017,Nojiri:2006ri}. 
In brief, there are two possible approaches to describe all observational datasets from 
planetary to cosmological scales: the first is to preserve GR by adding extra particles and/or
scalar fields; the second is to modify the geometrical description of the space-time. 
Both must be tested in all possible astronomical scenarios in order to understand at which scales
their contributions become significant. Let us note that some of these modified 
theories have been ruled out using the recent discovery of the electromagnetic counterpart associated 
to the emission of the gravitational waves \cite{Ligo2017,felix,Bellucci:2008jt,graviton,Bogdanos:2009tn, Ezquiaga2017, Sakstein2017,Lombriser2016}.  
Such a discovery opens new avenues to test modified theories of gravity further, and those tracks must be explored.

The simplest prescription
to modify GR is to generalize the Einstein-Hilbert Lagrangian to an arbitrary function of the Ricci scalar, $f(R)$. 
Then, one should take care of the fact that, in the weak field limit, any alternative relativistic theory of gravity 
must reproduce GR in order to recover the tight constraints at the Solar System scale \cite{Will93, Stairs2003, Everitt2011}. 
Here, we are interested in the post-Newtonian limit to describe the motion of test-particles (and more in general, of a system). 
In models where unknown particles/scalar fields are added to GR, in order to recover the Solar System bounds,
one has to require that such scalar fields are screened in a high density environment. However, these mechanisms are imposed {\em ad-hoc}
to avoid that scalar fields dominate the dynamics of small scale systems. In the case of $f(R)$-gravity the gravitational potential 
is modified by a Yukawa-like term related to a new characteristic scale length of the system that appears 
because one has to solve forth (instead of second), order field equations, and this new scale length can act automatically 
as a screening mechanism \cite{Annalen}. 

Some of the most promising objects to test the underlying theory of gravity are pulsars. These objects are very dense and 
rapidly rotating (up to hundred times per second) neutron stars emitting gamma radiation beams or X-rays.
They act as a very precise clock and any deviation in their pulse from the one predicted by GR can be detected.
These deviations can be related to the violation of the strong equivalence principle and the 
variation of the gravitational constant. Both circumstances have been investigated using binary systems composed 
by a pulsar and another massive object (such as a neutron star or a white dwarf) 
that produces these anomalies in the pulse \cite{pulsarbook}. Anyway, these deviations can also 
be interpreted as a signature of an alternative theory of gravity \cite{deLa2012,deLa_deMa2014,deLa_deMa2015,Berry2011,LeeS2017,Liu2018}.
Forthcoming observations will increase the current point source sensitivity and resolution by combining different
facilities such as large telescopes apertures, adaptive optics, and near infrared (NIR) interferometry, and they will allow to detect 
pulsars with orbital period in scales as low as one year. Therefore, the measure of the periastron shift will became one of the most promising
tools to test GR and alternative theories \cite{Berti,Antoniadis,Freire,Shao}.
The most rigorous test of  alternative theories would be provided by a pulsar orbiting near a  
supermassive black hole (SMBH) \cite{Liu2012}. In such a case, we would not only expect the largest deviations from GR, but we could also measure the properties of the Black Hole (BH). 
A pulsar-BH system has not been found yet, but the prospects of finding  one such can increase
enormously within the curved space-time around Sagittarius A* (Sgr A*), 
the SMBH at the center of the Milky Way \cite{Goddi2017,Angelil2010}.

In order to be measurable with current instruments, pulsars with short orbital periods would need to be discovered,  
such pulsars would orbit at distances inside a $10$ AU radius circle centered at Sgr A*. 
In particular, an ideal pulsar would be one 
spinning a few hundred times per second. Searches are currently undergoing with the BlackHoleCam\footnote{https://blackholecam.org} 
and Event Horizon Telescope (EHT) Collaboration\footnote{http://www.eventhorizontelescope.org} \cite{Goddi2017,Akiyama2015,Huang2007,Doeleman2008,Doeleman2009,Doeleman2012,Falcke2000,Fish2011,Fish2016,Johnson2015,Younsi2016,Abdujabbarov2015}. 
EHT is a project to create a large telescope array consisting of a global network of radio telescopes and combining data from several Very-Long-Baseline Interferometry (VLBI) stations around the Earth. 
The aim is to observe the immediate environment of the Galactic Center, as well as the even larger BH in 
Messier $87$ (M$87$), with angular resolution comparable to the BH's event horizon \cite{Doeleman2012}.
These facilities, together with current and forthcoming Pulsar Timing Array (PTA) observatories \cite{Hobbs2012}, will give us a unique
opportunity to test alternative theories of gravity using the orbital motion of a test particle around a massive object as well 
as the motion of a binary system. Hereby, we are indeed currently building
the theoretical facilities needed to test $f(R)$-gravity. 

The aim is to demonstrate the capability of the Yukawa-like gravitational potential of explaining the dynamics of the particles at the Galactic Center.  
The study of the periastron shift is complementary to other studies
on the time variation of the orbital period in $f(R)$ gravity that have been used to constrain the graviton mass
\cite{deLa_deMa2014, deLa_deMa2015}. Although the periastron shift has been studied in a sort of
semi-classical approach where the Yukawa-potential has been considered to describe the gravitational force in the Newtonian classical dynamics \cite{Borka20012, Borka20013, Zakharov2016, idmRLmdl2017,Iorio}, the full relativistic approach is needed to take into 
account the geodesic structure of the space-time, and to investigate how  particles dynamics is affected. 
The systems  that we will examine are  somewhat idealized, compared to real astrophysical sources. 
For example, we neglect tidal effects that become important only when the mean separation of 
the two objects is of the order of their radius. This allow us to understand the essence of the physical mechanism 
with minimal complications, and to form the basis for a more detailed study of realistic sources in alternative theories of gravity.
The paper is divided as follows: in Sect. \ref{due} we briefly review the post-Newtonian limit of an analytic $f(R)$ model showing 
how the Yukawa-like gravitational potential arises; in Sect. \ref{tre}, we introduce the geodesic motion in $f(R)$ gravity computing the 
geodesic equation and the canonical momenta; in Sect. \ref{quattro}, we solve numerically the geodesic equation illustrating the effect of the
Yukawa-potential on the orbital precession; in Sect. \ref{cinque}, we compute an analytic formula for the periastron advance and apply it
to toy models; finally in Sect. \ref{sei}, we give our conclusion and remarks.

\section{Post-Newtonian limit and Yukawa-like gravitational potentials}
\label{due}
Here we summarize the main steps that lead to the modification of the gravitational potential 
in the post-Newtonian limit of the $f(R)$-gravity. The natural starting point is to
consider a general fourth order gravity action:
\begin{equation}\label{actfR}
\mathcal{A}\, = \,\int
d^4x\sqrt{-g}\biggl[f(R)+\mathcal{X}\mathcal{L}_m\biggr]\,,
\end{equation}
where $f(R)$ is an  analytic function of Ricci scalar, $g$ is the
determinant of the metric $g_{\mu\nu}$, $\mathcal{X}=16\pi
G/c^4$ is the coupling constant and $\mathcal{L}_m$ describes
the standard fluid-matter Lagrangian. For $f(R)=R$,
the Hilbert-Einstein action of GR is restored.

Varying the action in Eq. \eqref{actfR} with respect to the metric tensor we obtain the following field equations:
\begin{equation}\label{fe}
f'(R)R_{\mu\nu}-\frac{1}{2}f(R)g_{\mu\nu}-f'(R)_{;\mu\nu}+g_{\mu\nu}\Box
f'(R)=\frac{\mathcal{X}}{2}T_{\mu\nu}\,,\\
\end{equation}
and their trace
\begin{equation}\label{fetr}
3\Box f'(R)+f'(R)R-2f(R)=\frac{\mathcal{X}}{2}T\,.
\end{equation}

Here primes indicate derivatives with respect to the Ricci curvature, and $\Box$ is the usual d'Alembert operator.
{The next step is the fairly common practice to make a conformal transformation 
to pass from the Jordan frame to the Einstein frame, in which the field equations are reduced from 
fourth order partial differential equations to second order ones, and a scalar field arises from the extra degrees of freedom.
On the one hand, this operation simplifies the calculations and requires to introduce a mechanism to screen the scalar field in high 
density environments (short distances) \cite{Khoury2004,Khoury2009,defelice2010}. 
On the other hand, the two frame are mathematically equivalent but 
their physical equivalence is, nowadays, under debate \cite{darkmetric,Magnano,Faraoni}. To be sure of the physical equivalence 
one should reproduce the results in both frames and compare them. The alternative is to stay in Jordan frame accepting the idea 
of having to handle with the fourth order field equations in Eq. \eqref{fe}, and regarding to the extra degrees of freedom of the theory as free 
parameters to be constrained with the data. This approach avoids the need of introducing a screening mechanism because of the scale
dependence of the theory. Thus, hereafter, all calculations will be performed in the Jordan frame.}

Following \cite{arturosferi,arturonoether}, the post-Newtonian (PN) limit of $f(R)$ gravity 
can be computed assuming a general  spherically symmetric metric\,:
\begin{eqnarray}\label{me}
ds^2\,&&=\,g_{tt}(x^0,r)d{x^0}^2-g_{rr}(x^0,r)dr^2-r^2d\Omega^2\,,
\end{eqnarray}
where $x^0\,=\,ct$ and $d\Omega^2$ is the solid angle. For the sake of simplicity, following \cite{Annalen}
we set $c=1$ (it will be restored in the next sections).
Then, let us add perturbations of the metric tensor with respect to a Minkowskian background
$g_{\mu\nu}\,=\,\eta_{\mu\nu}+h_{\mu\nu}$, and  assume
an $f(R)$ Lagrangian expandable in Taylor series:
\begin{eqnarray}\label{sertay}
f(R)&&=\sum_{n}\frac{f^n(R_0)}{n!}(R-R_0)^n\nonumber\\&&\simeq
f_0+f'_0R+f''_0R^2+f'''_0R^3+...\,.
\end{eqnarray}

Inserting the Eq.~\eqref{sertay} into 
field equations \eqref{fe} - \eqref{fetr} and expanding them
up to orders ${\mathcal O}(0)$, ${\mathcal O}(2)$ and ${\mathcal O}(4)$, one obtains
\begin{eqnarray}\label{eq2}
&&f'_0rR^{(2)}-2f'_0g^{(2)}_{tt,r}+8f''_0R^{(2)}_{,r}-f'_0rg^{(2)}_{tt,rr}+4f''_0rR^{(2)}=0\,,\nonumber\\
&&f'_0rR^{(2)}-2f'_0g^{(2)}_{rr,r}+8f''_0R^{(2)}_{,r}-f'_0rg^{(2)}_{tt,rr}=0\,,\nonumber\\
&&2f'_0g^{(2)}_{rr}-r\left[f'_0rR^{(2)}-f'_0g^{(2)}_{tt,r}-f'_0g^{(2)}_{rr,r}+4f''_0R^{(2)}_{,r}+\right.\nonumber\\&&\left.+4f''_0rR^{(2)}_{,rr}\right]=0\,,\nonumber\\
&&f'_0rR^{(2)}+6f''_0\left[2R^{(2)}_{,r}+rR^{(2)}_{,rr}\right]=0\,,\nonumber\\
&&2g^{(2)}_{rr}+r\left[2g^{(2)}_{tt,r}-rR^{(2)}+2g^{(2)}_{rr,r}+rg^{(2)}_{tt,rr}\right]=0\,.
\end{eqnarray}
Using the trace equation (the fourth in system \eqref{eq2}), one gets the following general solution:
\begin{align}
 \label{sol} g^{(2)}_{tt}&=\delta_0-\frac{\delta_1}{f'_0 r}+\frac{\delta_2(t)\lambda^2 e^{-r/\lambda}}{3}+\frac{\delta_3(t)\lambda^3 e^{r/\lambda}}{6r}\,,\\
 \label{sol1} g^{(2)}_{rr}&=-\frac{\delta_1}{f'_0r}-\frac{\delta_2(t)\lambda^2(1+r/\lambda)e^{-r/\lambda }}{3r}\nonumber\\
&+\frac{\delta_3(t)\lambda^3(1-r/\lambda)e^{r/\lambda }}{6r}\,,\\
 \label{sol2} R^{(2)}&=\delta_2(t)\frac{e^{-r/\lambda}}{r}+\frac{\delta_3(t)\lambda e^{r/\lambda}}{2r}\,,
\end{align}
where $\lambda\,\doteq\,\sqrt{-6f''_0/f'_0}$, the constant $\delta_0$ can be neglected,
the $\delta_1$ is an arbitrary constant, and  $\delta_2(t)$ and $\delta_3(t)$
are completely arbitrary functions of time which, 
since the differential equations in the system (\ref{eq2}) contain only spatial
derivatives, can be fixed to  constant values. 
Let us note that on
the limit $f(R)\rightarrow R$, for a point-like mass $M$, we recover the standard
weak field limit when $\delta_1 =GM$. 
Finally, requiring that the metric must be asymptotically flat 
(Yukawa growing mode  in the system of Eqs. (\ref{sol})-(\ref{sol2}) are discarded)
one obtains
\begin{align}
\label{mesol} g_{tt}(x^0,r)&=\,1-\frac{GM}{f'_0r}+\frac{\delta_2(t)\lambda^2 e^{-r/\lambda}}{3}\,,\\
\label{mesol1} g_{rr}(x^0,r) &= 1+\frac{GM}{f'_0r}+\frac{\delta_2(t)\lambda^2(1+r/\lambda)e^{-r/\lambda }}{3r}\,,\\
R\,&=\,\frac{\delta_2(t)e^{-r/\lambda}}{r}\,.
\end{align}

The metric in Eqs.~\eqref{mesol} and ~\eqref{mesol1} also contains the solution of the modified gravitational potential.
Specifically, remembering that $g_{00}\,=\,1+2\Phi_{grav}\,=\,1+g_{tt}^{(2)}$ \cite{Weinberg1972}, one can extract the expression
for the gravitational potential in $f(R)$-gravity:
\begin{eqnarray}\label{gravpot}
\Phi\,=\,-\frac{GM}{f'_0r}+\frac{\delta_2(t)\lambda^2e^{-r/\lambda}}{6r}\,.
\end{eqnarray}
Let notice that the standard Newtonian potential is
recovered only in the particular case $f(R)=R$ while it is not so
for generic analytic $f(R)$ models.  Eq.~\eqref{gravpot} can be 
straightforwardly recast as (for more details see \cite{PhysRept, Annalen})
\begin{equation}
\label{gravpot1} \Phi(r) = -\frac{G M}{
(1+\delta) r}\left(1+\delta e^{-\frac{r}{\lambda}}\right)\,,
\end{equation}
by defining $\displaystyle{1+\delta=f'_0}$, and assuming that
$\delta_1$ is quasi-constant, and it is related to $\delta$ as follows
through
\begin{equation}
\delta_2=-\frac{6GM}{\lambda^2}\frac{\delta}{1+\delta}\,.
\label{eq:delta1}
\end{equation}

Eq.~(\ref{gravpot1}) deserves some comments. If $\delta=0$ then the Newtonian potential is recovered.
Next, the first term is the Newtonian  potential generated by a point-like mass $\displaystyle{\frac{M}{1+\delta}}$. 
And, the second term is the Yukawa-like modification of the gravitational potential with a scale
length, $\lambda$, related to the above coefficient of the Taylor expansion of the gravitational 
Lagrangian. The parameter $\lambda$ naturally arises from the theory, and acts as a screening mechanism. 
It makes the Yukawa correction be negligible at small scales while 
relevant at galactic, extragalactic and cosmological scales providing a possible way to
explain galaxy rotation curves, cluster of galaxies and the accelerated expansion of the Universe
without requiring Dark Matter and/or Dark Energy \cite{ demartino2014, darkmetric,idm2015, demartino2016, Cardone2011, Napolitano2012, Cap-def-Sal2009}.

Understanding  the amplitude of these corrections  to the gravitational potential at the scale of the 
stellar systems is one of the most important tools that could be used to observationally confirm or rule out 
these alternative approaches to GR.

\section{Geodesic motion in f(R)-gravity}
\label{tre}

\begin{figure}[!ht]
 \includegraphics[width=\columnwidth]{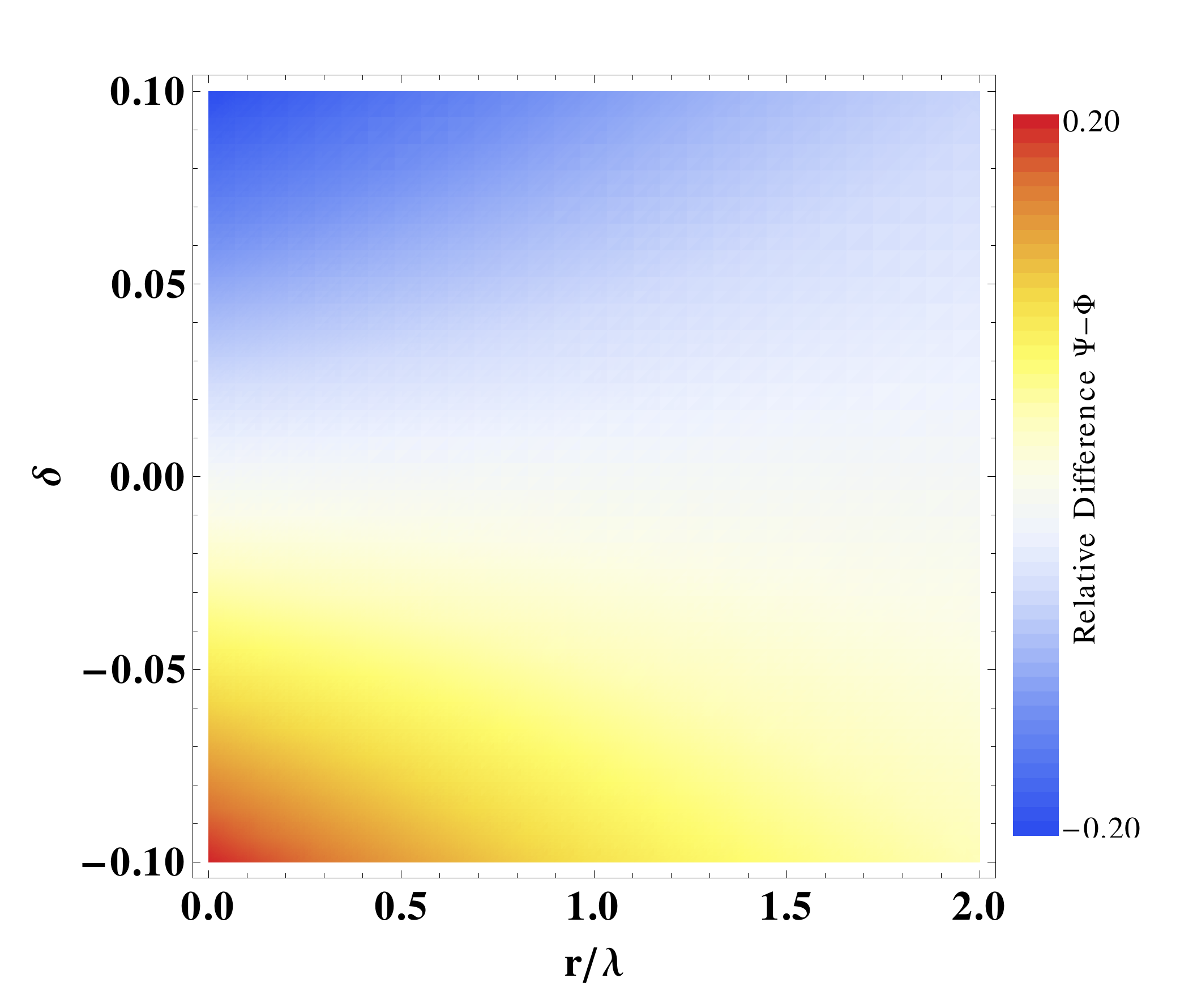}
 \caption{Relative difference of the two gravitational fields $\Phi(r)$ and $\Psi(r)$ as a function of the parameters of the 
 strength and the scale length of the Yukawa term in Eq.~\eqref{gravpot1}. Here, we have used $G=M=c=1$.}\label{fig:psi_phi}
\end{figure}

Let's apply the Euler-Lagrange equations to find the  geodesics equations of motion associated to the line element given in Eqs. 
\eqref{mesol} and \eqref{mesol1}. After some manipulations, they can be recast into the following form
\begin{equation}
\label{metric0}
ds^2=\left[1+\Phi(r)\right]dt^2-\left[1-\Psi(r)\right]dr^2-r^2d\Omega\,,
\end{equation}
where the two potentials $\Phi(r)$ and $\Psi(r)$ are given by
\begin{eqnarray}
\label{eq:PHI}\Phi(r) &=& -\frac{2G M \left(\delta  e^{-\frac{r}{\lambda}}+1\right)}{rc^2(\delta +1)}, \\
\label{eq:PSI}\Psi(r) &=& \frac{2G M }{rc^2}\biggl[\frac{\left(\delta  e^{-\frac{r}{\lambda}}+1\right)}{(\delta +1)}+\frac{\left(\frac{\delta  r
   e^{-\frac{r}{\lambda}}}{\lambda}-2\right)}{(\delta +1)}\biggr]\,, 
\end{eqnarray}
with the speed of light having been reinstated. Note that the potential $\Psi(r)$  can be rewritten as
\begin{equation}
\Psi(r) = \Phi(r) + \delta\Phi(r)\,, 
\end{equation}
where the term $\delta\Phi(r)$ representing an extra contribution to the total
gravitational potential. Since we are interested in small scale systems\footnote{Here "small scales" means stellar system scales.}, we have verified whether such contribution is negligible or not. 
In  Fig.~\ref{fig:psi_phi}, we show the region plot of the ratio 
$(\Psi(r) -\Phi(r))/\Phi(r)$.  
Since such ratio is almost insensitive to the scale length $\lambda$, the latter has been kept fixed to the confidence value of 
$5000$ AU \cite{Borka20012,Borka20013,Zakharov2016,Zakharov2018}.
The color bar on the figure indicates the relative change of the two potentials.
We have varied $\delta$ from $-0.1$ to $0.1$ showing that
the departure of $\Psi(r)$ from $\Phi(r)$ is $\sim20\%$ for $\delta=\pm0.1$, while 
it decreases to  $\sim2\%$ for $\delta\sim\pm0.01$. 
To explain binary systems in the framework of $f(R)$ gravity, 
we need  very small departure from GR, which means $|\delta|\ll 0.1$ \cite{deLa_deMa2014, deLa_deMa2015}. 
Thus,  hereafter, we will assume $\Psi(r)\sim\Phi(r)$.

Thus, the line element in Eq.~\eqref{metric0} becomes
\begin{equation}
\label{metric}
ds^2=\left[1+\Phi(r)\right]dt^2-\left[1-\Phi(r)\right]dr^2-r^2d\Omega.
\end{equation}
To compute the geodesic equations, we use the Euler-Lagrange equations:
\begin{equation}
\frac{d}{ds}\frac{\partial \cal L}{\partial \dot x^\mu}-\frac{\partial \cal L}{\partial x^\mu}=0\,,
\end{equation}
that are equivalent to the geodesic equations \cite{Weinberg1972}
\begin{eqnarray}
\ddot{x}^\mu + \Gamma^{\mu}_{\alpha\beta}\dot{x}^\alpha\dot{x}^\beta=0\,.
\end{eqnarray}
For the line element in Eq. \eqref{metric}, the non-zero Levi-Civita connections are
\begin{align}
 \Gamma^1_{1 1} & = -\frac{R_S \left[\left(e^{\frac{r}{\lambda }}+\delta \right) \lambda +\delta  r\right]}{\lambda  r \left[2 R_S \left(e^{\frac{r}{\lambda }}+\delta \right)+e^{\frac{r}{\lambda }} (1+\delta ) r\right]}\,, \\
 \Gamma^1_{2 2} & =-\frac{e^{\frac{r}{\lambda }} (1+\delta ) r^2}{2 R_S \left(e^{\frac{r}{\lambda }}+\delta \right)+ e^{\frac{r}{\lambda }} (1+\delta ) r}\,, \\
 \Gamma^1_{3 3} & =-\frac{e^{\frac{r}{\lambda }} (1+\delta ) r^2 \sin^2\theta}{2 R_S \left(e^{\frac{r}{\lambda }}+\delta \right)+ e^{\frac{r}{\lambda }} (1+\delta ) r}\,, \\
 \Gamma^1_{0 0} & =\frac{R_S \left[\left(e^{\frac{r}{\lambda }}+\delta \right) \lambda +\delta  r\right]}{\lambda  r \left[2 R_S\left(e^{\frac{r}{\lambda }}+\delta \right)+ e^{\frac{r}{\lambda }} (1+\delta ) r\right]}\,, 
 \end{align}
 \begin{align}
 \Gamma^2_{2 1} & =\frac{1}{r}\,, \\
 \Gamma^2_{3 3} & =-\cos\theta \sin\theta\,, \\
 \Gamma^3_{3 1} & =\frac{1}{r}\,, \\
 \Gamma^3_{3 2} & =\cot\theta\,, \\
 \Gamma^0_{0 1} & =\frac{R_S\left[\left(e^{\frac{r}{\lambda }}+\delta \right) \lambda +\delta  r\right]}{\lambda  r \left[e^{\frac{r}{\lambda }} (1+\delta ) r -2 R_S \left(e^{\frac{r}{\lambda }}+\delta \right)\right]}\,.
\end{align}
 Here, we have introduced the definition of the general relativistic Schwarzschild radius: $R_S = GM/c^2$ 
 and we have eliminated the proper time. Finally, the geodesics equations are:
\begin{align}
{\ddot r} =&\Delta^{-1}\biggl[R_S \left({\dot r}^2-{\dot t}^2\right) \left(\delta  (\lambda +r)+ e^{\frac{r}{\lambda }} \lambda\right)\nonumber\\
&+e^{\frac{r}{\lambda }} \lambda (1+\delta ) r^3 \left({\dot \theta}^2+\sin^2\theta {\dot \phi}^2\right)\biggr]\,, \label{eq:geo1} \\[0.2cm]
\label{eq:geo2} {\ddot \theta} =&\cos\theta \sin\theta {\dot \phi}^2 -  \frac{2 {\dot r} {\dot\theta}}{r}\,,\\[0.2cm]
\label{eq:geo3}  {\ddot \phi} =& -\frac{2 {\dot \phi}}{r} \left[{\dot r}+\cot\theta r {\dot \theta}\right]\,,\\[0.2cm]
\label{eq:geo4}  {\ddot t} =& \Delta^{-1}\biggl[2 R_S \left[\left(e^{\frac{r}{\lambda }}+\delta \right) \lambda +\delta  r\right] {\dot r} {\dot t}\biggr]\,,
\end{align}
where, for the sake of convenience, we have defined
\begin{equation}
 \Delta\equiv\lambda  r \left[2 R_S \delta +e^{\frac{r}{\lambda }} \left(2 R_S- (1+\delta ) r\right)\right]\,.
\end{equation}
The above equations can be integrated numerically  to obtain the orbital motion and precession of a two-body system.
Although this represents a powerful tool to study the orbital motion of the stars around a massive object, such as the 
S-stars around the SMBH at the center of the Milky way galaxy, an analytical solution to predict the periastron advance would 
be more convenient for studies of binary systems of neutron stars and/or white dwarfs. To this aim, we must
define the Lagrangian associated to the metric elements of Eq. \eqref{metric}
\begin{eqnarray}
2{\cal L}=\left[1+\Phi(r)\right]{\dot t}^2-\left[1-\Phi(r)\right]{\dot r}^2-r^2{\dot \theta}^2-r^2\sin^2\theta {\dot\phi}^2\,. \nonumber\\
\label{lagrangianR}
\end{eqnarray}
Then, the canonical momenta are 
\begin{eqnarray}
p_t&\equiv&\frac{\partial{\cal L}}{\partial\dot{t}}=\left[1+\Phi(r)\right]\dot{t}\,,\\
p_r&\equiv&\frac{\partial{\cal L}}{\partial\dot{r}}=-\left[1-\Phi(r)\right]{\dot{r}}\,,\\
p_\theta&\equiv&\frac{\partial{\cal L}}{\partial\dot{\theta}}=-r^2\dot{\theta}\,,\\
p_\phi&\equiv&\frac{\partial{\cal L}}{\partial\dot{\phi}}=-r^2\sin^2\theta\dot{\phi}\,.
\end{eqnarray}
Next, if we write the Euler-Lagrange equations for the time component we obtain 
\begin{eqnarray}
\frac{d}{ds}\left[\left(1+\Phi(r)\right){\dot t}\right]=0\,.
\end{eqnarray}
The latter implies there is a conserved quantity we will call energy: 
\begin{eqnarray}
p_t\equiv\left[1+\Phi(r)\right]{\dot t}\equiv E\,.
\label{Energy}
\end{eqnarray}
Then, we find the $\phi$ component of the Euler-Lagrange equation
\begin{eqnarray}
\frac{d}{ds}\frac{\partial \cal L}{\partial {\dot \phi}}=\frac{\partial \cal L}{\partial \phi}=0\,,
\end{eqnarray}
which also leads us to define a conserved quantity:
\begin{eqnarray}
p_\phi\equiv r^2\sin^2\theta\dot{\phi}\equiv L\,,
\label{AngularM}
\end{eqnarray}
where $L$ is the angular momentum per unit mass of the two bodies.
From the equation for the $\theta$ component  we find
\begin{eqnarray}
\frac{d}{ds}\frac{\partial {\cal L}}{\partial{\dot \theta}}=\frac{\partial {\cal L}}{\partial{\theta}}\neq0\,,
\end{eqnarray}
which is not a conserved quantity. Thus, the $\theta$ equation reads:
\begin{eqnarray}
\frac{d}{ds}(r^2{\dot \theta})=r^2{\dot \phi}^2\sin \theta \cos \theta.
\end{eqnarray}
Finally we need to compute the $r$ equation, which is quite involved because of  the heavy explicit dependence on $r$
in the metric.

Since we want to study the orbits, as a first step we may simplify the problem by using its symmetries. 
Therefore, we fix the coordinate system so that 
the orbit of the particle lies on the plane ($r-\phi$), and fix the $\theta$ coordinate to be $\pi/2$ so that 
$\dot \theta=0$. Since we are interested on studying only time-like geodesics \cite{Chandrasekhar1983}, we use 
the constants of motion defined in the above equations to obtain the following identity:
\begin{eqnarray}
&&E^2\left[1+ \Phi(r)\right]^{-1}-\frac{L^2}{r^2}-\frac{\left[\Phi(r)-1\right]^2 }{1-\Phi(r)}{\dot r}^2=1\,.
\label{lag}
\end{eqnarray} 
Finally, by solving Eq. \eqref{lag},  we get an explicit equation for ${\dot r}^2$:
\begin{eqnarray}
{\dot r}^2=\frac{L^2 \left[\Phi(r)+1\right]-E^2 r^2}{r^2
   \left[\Phi(r)-1\right] \left[\Phi(r)+1\right]}\,.
   \label{rdot2}
 \end{eqnarray}
The equations we have built are needed to compute the periastron shift discussed/calculated in Sect. \ref{cinque}. 

 \section{Numerical Solutions of the geodesic equations of motion}
 \label{quattro} 

In order to show how the Yukawa correction to the Newtonian 
potential affects the orbital motion, we solve  numerically
the geodesic equations \eqref{eq:geo1}-\eqref{eq:geo4}.
Those parametric differential equations are non-linear, thus,
in order to have a well-posed Cauchy problem, we have to define
the initial and boundary conditions. We solve the Cauchy problem, 
as in the classical case, with the initial conditions 
$\dot{r}(\tau=0)= 0$, $\dot{\phi}(\tau=0)= 0$, $\dot{\theta}(\tau=0)= 0$, and 
$\theta(\tau=0)=\pi/2 $, obtaining a not planar solution ($\ddot{\theta}\ne 0$).
We perform a set of simulations varying the parameters 
$\{ M, E, \epsilon, \dot{r}(\tau=0), \dot{\phi}(\tau=0), \dot{\theta}(\tau=0)\}$ 
to account for the high non-linearity of the geodesic equations, and  
to obtain a set of parameters that guarantee the stability of the solution.
Once the numerical integration of the geodesic equation has been optimized, 
we are able to highlight the specific contributions of Yukawa-correction 
to orbital motion. 
Usually, one uses the orbital motion and the pulsar timing to study the properties of the SMBH at the center of the 
Milky Way (for detailed explanations on pulsar timing and other pulsar observing 
techniques see \cite{pulsarbook}). Here, we are going to use an inverse approach. 
We fix {\it a priori} the parameters of the SMBH to study the orbital motion of a pulsar-like object.
Specifically, we consider the SMBH at the center of the Milky Way galaxy, SgrA*, having a mass $M=(4.5\pm0.6)\times 10^6M_{\odot}$ 
\cite{Ghez2008} and located at a distance of $R_0\sim8$kpc from the Sun \cite{Eisenhauer2003}.
For convenience, we have fixed the scale length $\lambda=5000 AU$ \cite{Zakharov2016}, and set $G=c=1$. Thus, all results in the 
figures are given in physical units.

\begin{figure}[!ht]
 \includegraphics[width=0.99\columnwidth]{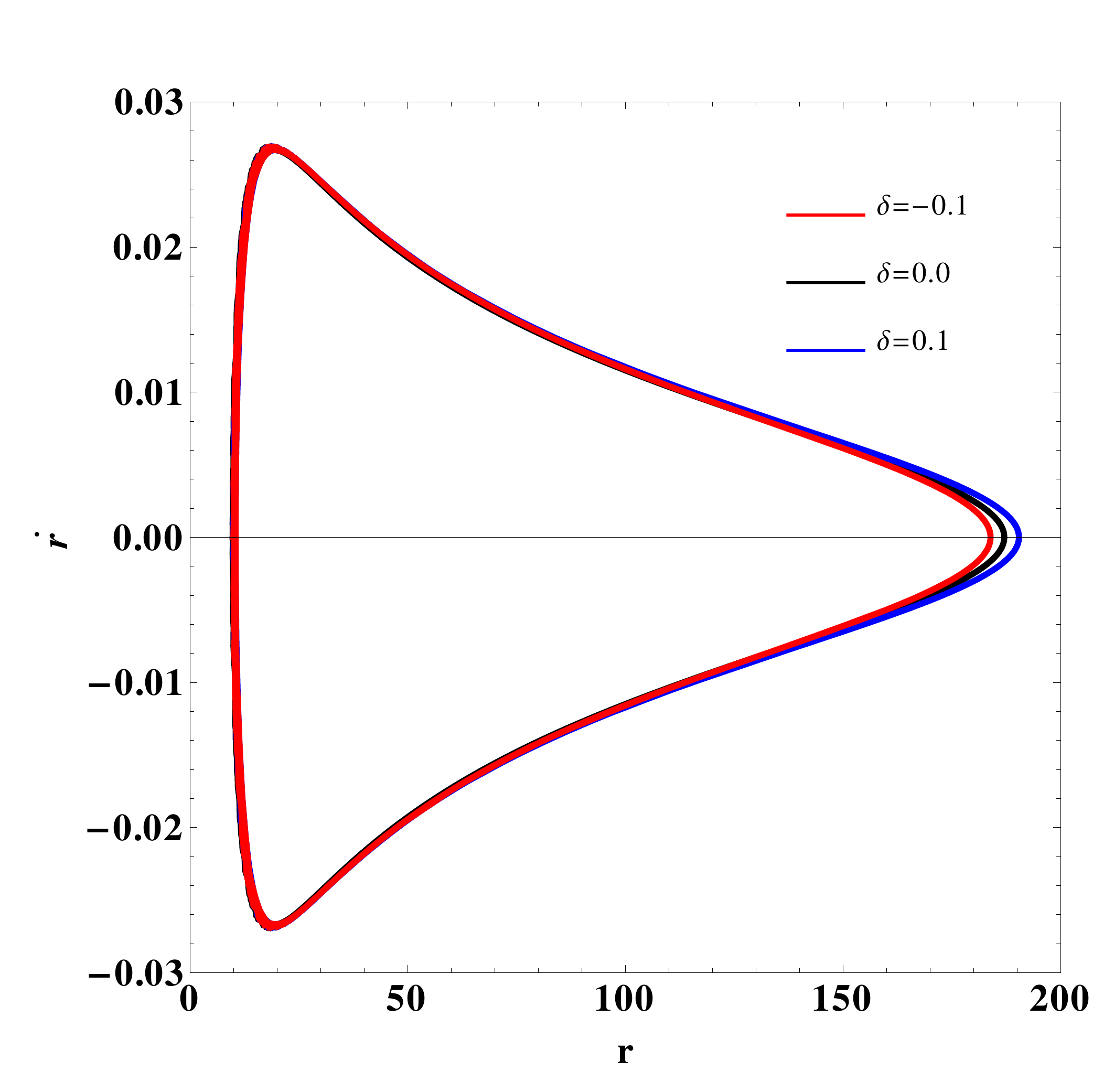}
 \caption{Phase space diagram of a closed orbit in the Yukawa potential.}\label{fig:rdot_r}
\end{figure}

In Fig. \ref{fig:rdot_r}, we illustrate the phase
portrait of $\dot{r}(\tau)$ versus $r(\tau)$ for the GR solution ($\delta=0$, black line), 
and for $\delta=-0.1$ and $\delta=0.1$ shown in red and blue lines, respectively. 
For both values of the $\delta$ the orbit assumes a stable configuration and, the Yukawa correction term 
induces departures form the configuration of the orbits obtained in GR. Specifically, for $\delta = -0.1$, 
the semi major axis is shorter,
 while for $\delta = 0.1$ is longer, than the GR one ($\delta = 0.1$).

\begin{figure}[!ht]
 \includegraphics[width=0.99\columnwidth]{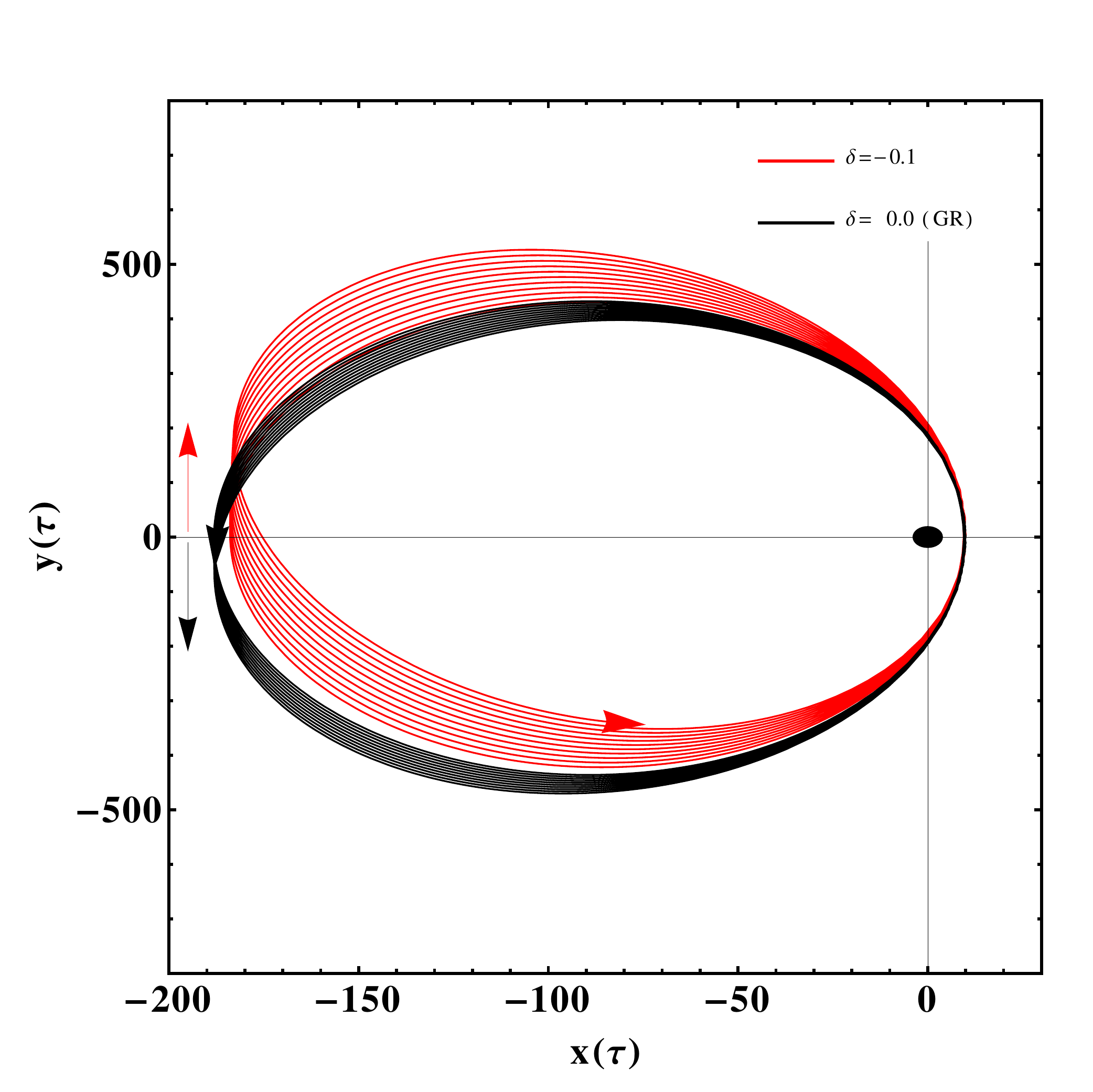}
 \caption{Numerical solution of the geodesic equation illustrating the periastron advance in the Yukawa-potential. 
 Here, we compare the GR solution ($\delta = 0$) and the one for $\delta = -0.1$.  
 The black dot point indicates the central object.}\label{fig:prec1}
\end{figure}

\begin{figure}[!ht]
 \includegraphics[width=0.99\columnwidth]{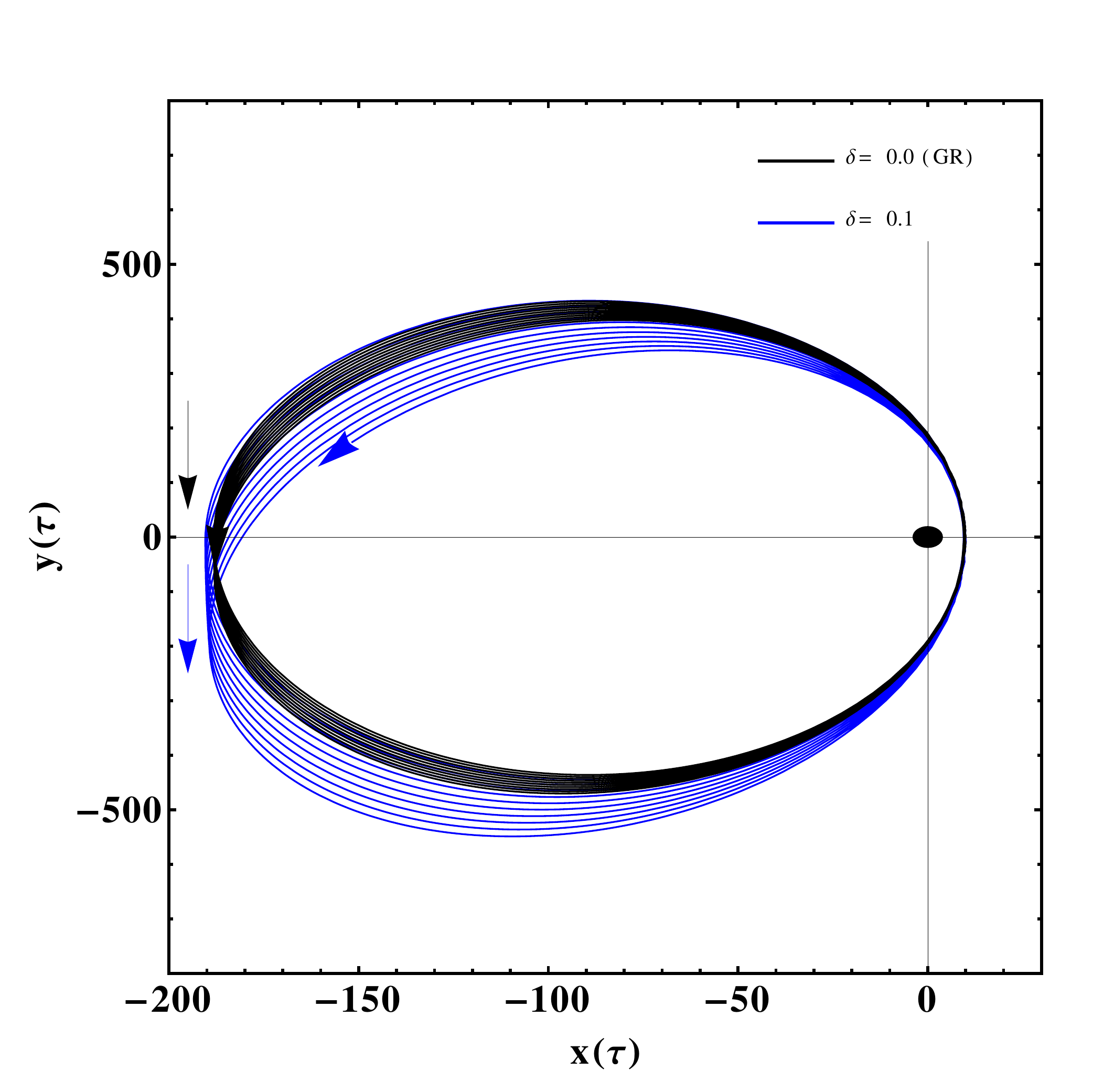}
 \caption{The plot follows the conventions adopted for Fig. \ref{fig:prec1}, while comparing the GR 
 with the Yukawa solution for $\delta = 0.1$.}\label{fig:prec2}
\end{figure}

{ The orbital precession is easily discernible drawing orbits. Thus, 
in Fig. \ref{fig:prec1} and \ref{fig:prec2}, we illustrate the periastron advance for both 
$\delta=-0.1$ and $\delta=0.1$  with a comparison with the general relativistic one. 
Let us note that the effect of the Yukawa-term is always to enhance the orbital precession
while its sign can change from being positive ($\delta>0$) to being negative ($\delta<0$). 
The numerical integration of geodesic equations qualitatively confirms
previous results found in semi-classical approaches  \cite{Borka20012, Borka20013, Zakharov2016, idmRLmdl2017}. 
This effect is due to the exponential term in the gravitational potential and it is negligible in binary systems. 
Nevertheless, it becomes viewable when  simulating an object  orbiting around a SMBH on scales comparable with $\lambda$, and 
it can be used to reduce further the parameter space of $f(R)$ gravity as previously suggested 
by \cite{Borka20012, Borka20013, Zakharov2016, idmRLmdl2017}.}

 
 \section{Periastron shift in Yukawa-like potential}
 \label{cinque}
 The most suitable candidates to test theories of gravity are binary systems constituted by a SMBH and an orbiting star \cite{DeLaurentis2018}. 
 Even just finding one normal pulsar around the BH will be phenomenally interesting to test alternative theories of gravity.
Generally speaking, an orbit closes if the angle $\phi$ sweeps out exactly $2\pi$ in the passage between two successive 
inner or two successive outer radial turning points. If the orbits precess,  $\phi$ changes by more 
than $2\pi$ between successive radial turning points.

To obtain an analytic formula for the periastron advance we need to  obtain the orbits $r=r(\phi)$.
 Thus, we replace the variable $\tau$ by $\phi$ with the aid of the angular momentum law
 Eq. \eqref{AngularM} and of Eq. \eqref{Energy}, and we obtain
\begin{equation} 
\left( \frac{dr}{d\phi}\right)^2=-\frac{r^2 \left[r^2 \left(\Phi(r)-E^2+1\right)+L^2 \left(\Phi(r)+1\right)\right]}{L^2
   \left[\Phi(r)^2-1\right]}\,,\nonumber\\
    \end{equation}
 that explicitly assumes the following form   
  \begin{eqnarray}    
\left( \frac{dr}{d\phi}\right)^2&=&\frac{2 \delta  G M e^{-\frac{r}{\lambda }}}{c^2 (\delta +1)L^2 r}+\frac{2 G M}{c^2 (\delta +1) L^2 r}+\frac{2 \delta  G M e^{-\frac{r}{\lambda }}}{c^2 (\delta +1) r^3}\nonumber\\&&+\frac{2 G M}{c^2 (\delta
   +1) r^3}+\frac{E^2}{L^2}-\frac{1}{L^2}-\frac{1}{r^2}\,.    
 \end{eqnarray}  
Let us perform the change of variable $u=1/r$, so that the previous equation reads  
\begin{eqnarray} 
\left( \frac{du}{d\phi}\right)^2=\frac{E^2-\left[\Phi(u)+1\right] \left[L^2
   u^2+1\right]}{L^2 u^4 \left[\Phi(u)^2-1\right]}\,.
 \label{dudphi} 
  \end{eqnarray}
 After some simplifications and imposing $\left( {du}/{d\phi}\right)^2=0$ we obtain
  \begin{eqnarray}\label{cicc}
\frac{2 \delta  G M u e^{-\frac{1}{\lambda  u}}}{c^2 (\delta +1)L^2}&+&\frac{2 GM u}{c^2 (\delta +1)L^2}+\frac{2 \delta  G M u^3 e^{-\frac{1}{\lambda 
   u}}}{c^2 (\delta +1)}\nonumber\\&+&\frac{2 G M u^3}{c^2 (\delta
   +1)}+\frac{E^2}{L^2}-\frac{1}{L^2}-u^2=0\,.\nonumber\\
     \end{eqnarray}
The most fruitful way to proceed is to rewrite 
 the previous equation in terms of orbital parameters. We introduce the eccentricity 
 $e$ and the {\it latus rectum} $l$ of the orbit, and we define the parameter  $\mu\equiv M/l$. 
 By definition, we use the ansatz that
  \begin{equation}\label{eq:uchi} 
  u=\frac{1+e\cos\chi}{l}\,,
     \end{equation}
    where $\chi$ is the so called {\it relativistic anomaly}. Thus, $\chi=0$ and $2\pi$ correspond to 
    successive periastron passages, and $\chi=\pi$ at intermediate apoastron. 
 Then, inserting Eq. \eqref{eq:uchi} in Eq. \eqref{cicc}, we obtain
\begin{eqnarray} \label{dchidphi}
\left( \frac{d\chi}{d\phi}\right)^2&&=\left[1-\left(e^2+3\right) \mu + 2\mu(e \cos\chi+1)^2 \right]\Upsilon
+\nonumber\\&& +\left(e^2-1\right) (1-4 \mu ) \mu ^2 -\mu ^2 (e \cos \chi+1)^2,
\end{eqnarray}
where we have defined the auxiliary variable
\begin{align}
& \Upsilon    = \frac{ 2 \mu ^2 (e \cos\chi +1) }{\delta+1}(\Upsilon_1+1),\,\,\\
& \Upsilon_1  =\delta  \biggl(\frac{1}{2 \lambda ^2 \mu ^2 (e \cos \chi+1)^2}-\frac{1}{\lambda  \mu  (e \cos \chi+1)}+1\biggr) \,.
\end{align}
Note that as we want to get an analytical solution and to study very 
close orbiting binary objects, we have expanded in Taylor series the exponential 
$e^{-\frac{1}{\lambda u}}$ up to the second order. Therefore, the use of the previous formula is restricted to the 
cases in which the semi-major axis of the orbit is much lower than the Yukawa scale length.
It is also important to note that when $\delta=0$ one recovers the well known results of GR \cite{Chandrasekhar1983}
  \begin{equation} 
  \left( \frac{d\chi}{d\phi}\right)^2=1-2\mu(3+e\cos \chi),\,\,
   \end{equation}   
that leads to the likewise well-known result
\begin{equation}\label{eq:deltaphiGR} 
 \Delta\phi_{GR}= \frac{6 \pi  G M}{a c^2 \left(1-e^2\right)}\,.
\end{equation}

The integration of Eq. \eqref{dchidphi} can be performed trivially, and  
finally it is possible to obtain the expression for the periastron advance
 \begin{eqnarray}\label{eq:deltaphi}   
\Delta\phi&=& \frac{\Delta\phi_{GR} }{(\delta +1)}\biggl( 1+
  \frac{2 \delta  G^2 M^2 }{3a^2 c^4  \left(1-e^2\right)^2}
  -\frac{2 \pi \delta  G^2 M^2 }{a c^4  \left(1-e^2\right) \lambda }  \nonumber\\&&-
  \frac{3\delta  G M}{a c^2  \left(1-e^2\right)}
     -\frac{ \delta G^2 M^2 }{6c^4(\delta +1) \lambda ^2}+
     \frac{\delta G M }{3 \lambda c^2}\biggr)\,.
  \end{eqnarray}

The Eq. \eqref{eq:deltaphi} shows explicitly that it reduces to Eq. \eqref{eq:deltaphiGR} for $\delta=0$. Next, 
the amount of relativistic precession depends by
\begin{itemize} 
  \item the values $M$ for the central mass,
  \item tight orbits (small values of $a$), 
  \item large eccentricities $e$,
  \item the Yukawa scale length $\lambda$,
  \item the Yukawa strength $\delta$.
  \end{itemize}
Therefore, the parameter space is larger than the one in the general relativistic case due to the presence of two 
extra parameters $\delta$ and $\lambda$ that affect the precession. As already mentioned in Sect. \ref{quattro}, 
the Yukawa-correction can change the sign of the precession as found in 
semi-classical approaches \cite{Borka20012, Borka20013, Zakharov2016, idmRLmdl2017, Zakharov2018}. 

 \subsection{Toy model stars around the Galactic center}
 \label{toymodel} 
\begin{figure}[!ht]
 \includegraphics[width=\columnwidth]{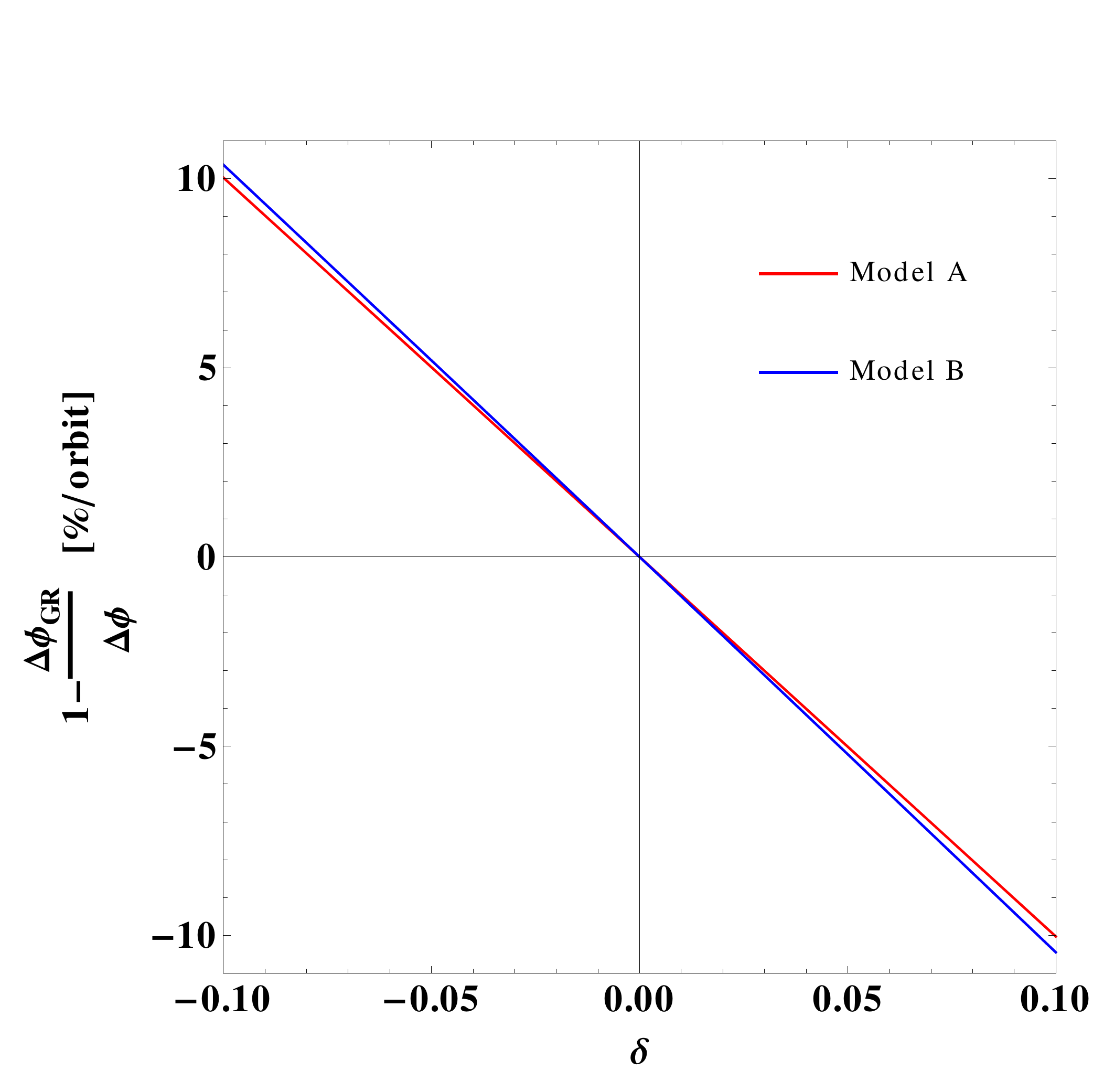}
 \caption{The plot illustrate the change of the periastron advance with respect to the GR one as function of $\delta$. 
 We used Eq. \eqref{eq:deltaphi} to compute analytically the periastron advance for a set of three toy  model stars orbiting
 around the Black Hole at the Galactic Center. The orbital parameters are given in Table. \ref{tab:1}.}\label{fig:deltaPHI_vs_delta}
\end{figure}

 Here we have particularized the periastron shift for a set of three toy model stars orbiting around the 
 BH in the Galactic Center. Let us remarks that being $\lambda\sim10^3$ AU we can not apply the 
 equation \eqref{eq:deltaphi} to the S-stars orbiting around the Galactic Center for which one should
 solve the geodesic equations numerically. The BH mass is fixed to $M_{BH}=4.5\times10^6 M_\odot$. 
 The orbital parameters of the three models are summarized in Table \ref{tab:1}. 
\begin{table}[!h]
\begin{center}
\caption{Values of periastron advance for different objects. 
In the table are reported the measured values of the eccentricity $e$, semi-major axis $a$ in meters, 
the general relativistic periastron advance, and the predicted values of$\Delta \phi$ for $\delta = \pm 0.01$ 
from Eq. \eqref{eq:deltaphi}.}
\label{tab:1}
\begin{tabular}{|lccccc|}
\hline\hline
Toy Model&  $e\quad$ & $a$ & $\Delta\phi_{GR}$ & $\Delta\phi_{\delta=-0.01}$ & $\Delta\phi_{\delta=0.01}$\\
         &           & ($10^{11}$m) & $(^\circ/orbit)$  & $(^\circ/orbit)$  & $(^\circ/orbit)$  \\
\hline
\hline
${\rm A}\quad$     & $0.678$  & $14.96$ & $8.88059$ & $8.97053$ & $8.79242$\\
${\rm B}\quad$     & $0.786$  & $7.48$  & $25.1087$ & $25.3642$ & $24.8583$\\
${\rm C}\quad$     & $0.888$  & $1.496$ &  $226.918$ & $229.303$ & $224.580$ \\
\hline
\end{tabular}
\end{center}
\end{table}
In Fig.~\ref{fig:deltaPHI_vs_delta}, we show the contribution of $f(R)$ gravity to the 
general relativistic periastron advance as a function of the strength of the Yukawa potential.
Here, the scale length has been fixed to the confidence value $\lambda=5000$ AU \cite{Zakharov2016}.
The figure shows that for $\delta>0$ the contribution of the Yukawa-correction increases the periastron
shift, while for $\delta<0$  it decreases it. The shift in the periastron advance in $f(R)$ gravity with respect to GR can 
reach an order of magnitude of $\sim10\%$ for $\delta=\pm0.1$, and it could be measurable with forthcoming observations 
of EHTC.

\begin{figure}[!ht]
 \includegraphics[width=\columnwidth]{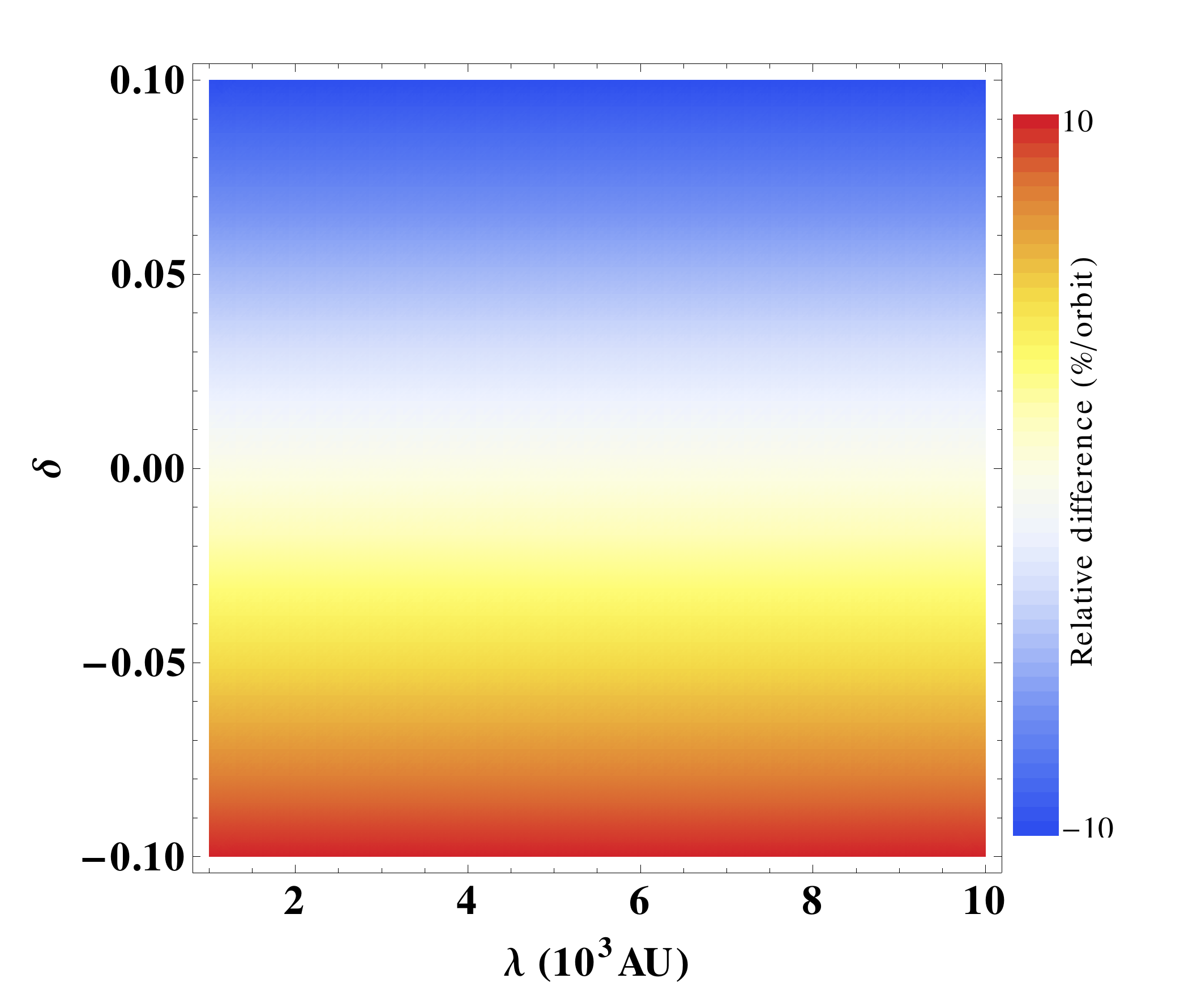}
 \caption{We illustrated the dependence of the periastron advance in Eq. \eqref{eq:deltaphi} from both the strength and the scale
 length of the Yukawa potential. The plot is particularized for the toy model A in Table \ref{tab:1}.}\label{fig:deltaPHI_vs_delta_vs_lambda}
\end{figure}

Finally, in Fig. \ref{fig:deltaPHI_vs_delta_vs_lambda}, we demonstrate that the impact of the scale length is
negligible, confirming the known degeneracy between $\delta-\lambda$ that cannot be constrained at the same time
using the orbital motion \cite{idmRLmdl2017}.

\subsection{Constraining Yukawa potential with pulsars in binary systems}

Binary systems composed by double pulsars or by a pulsar and a companion star provide a excellent 
laboratory to study alternative theories of gravity. It is well known that pulsars 
act as very precise clocks. Monitoring one such a clock allows us to measure 
the time of arrival (TAO) of pulses at the telescopes and to obtain the pulse profile.
In case the pulsar is part of a binary system, the pulse profile shows a periodic 
variation in the arrival time. This variation is related to the orbital motion around the
center of mass of the binary system, and it needs to be modeled. Binary systems can be described
in terms of the Keplerian parameters: the orbital period $P_b$, the projected semi-major axis 
$a_p \sin i$, the eccentricity of the orbit $e$, the periastron, $\omega$, and the time of the transition at
periastron $T_0$. Nevertheless, when considering close binary systems relativistic effects due to 
the strong field regime must be introduced. It is customary to parameterize 
the timing model using the post-Keplerian (PK) parameters: the time variation of the orbital period $\dot{P}_b$, 
the advance of the periastron $\dot{\omega}$, the time delay $\gamma$, and other two parameters, $r$ and $s$, 
related to the Shapiro delay due to the gravitational field of the companion star. 
Although GR is capable of describing those systems, alternative theories of gravity can be probed using 
specific generalizations of the PK parameters.  The main difference is that, in GR, the two masses
are the only free parameters. Therefore, observing two PK parameters leads to estimating the masses uniquely.
Clearly, precise measurements of the all PK parameters will provide an accurate estimation of the masses. 
Nevertheless, in $f(R)$-gravity this is not true.
The two masses are not the only free parameters, one also has the parameters of the gravitational potential ($\delta,\, \lambda$) 
or alternatively, their expression in terms of the Taylor coefficients ($f'_0,\, f''_0$), and they are degenerate with the masses. 
The only way to break this degeneracy is to
fix the masses \cite{deLa_deMa2014, deLa_deMa2015}. Therefore, calculating more PK parameters in alternative theories of gravity 
will give a powerful tool to estimate the masses of the two stars and, at same time, to constraint/rule our the theory.

The theoretical expression for the periastron advance in the case of binary systems is obviously dependent 
on the pulsar mass $m_p$ and on the mass of the companion star $m_c$. 
To generalize the periastron advance in Eq. \eqref{eq:deltaphi} 
to the case of a binary system we have to use Kepler's law and the fact that the total mass in Eq. \eqref{eq:deltaphi} 
can be recast as $M= m_c + m_p$. Thus, Eq.  \eqref{eq:deltaphi} becomes
\begin{eqnarray}  
\dot{\omega}&=& \frac{\dot{\omega}_{GR}}{(\delta +1)} \biggl[1
  +\frac{2 \delta}{ \left(1-e^2\right)^2}\biggl(\frac{2\pi}{P_b}\biggr)^{4/3} \frac{G^{4/3}}{c^4} (m_p+m_c)^{4/3} \nonumber\\&&
  -\frac{2\delta}{ \left(1-e^2\right) \lambda}\biggl(\frac{2\pi}{P_b}\biggr)^{2/3} \frac{G^{5/3}}{c^4} (m_p+m_c)^{2/3} \nonumber\\&&
  -\frac{2 \delta}{ \left(1-e^2\right)}\biggl(\frac{2\pi}{P_b}\biggr)^{2/3} \frac{G^{2/3}}{c^2} (m_p+m_c)^{2/3}\nonumber\\&&
  -\frac{\delta}{2 \lambda ^2} \frac{G^{2}}{c^4} (m_p+m_c)^2
  +\frac{\delta }{  \lambda }\frac{G}{c^2} (m_p+m_c)\,\biggr],
  \end{eqnarray}    
where the masses $m_p$ and $m_c$ are expressed in solar masses, and we have defined 
\begin{equation}
\dot{\omega}_{GR}= \biggl(\frac{2\pi}{P_b}\biggr)^{5/3} \frac{G^{2/3}}{c^2} \frac{(m_p+m_c)^{2/3}}{\left(1-e^2\right)} \,.
\end{equation}

The previous equation can be further simplified using 
the constant $ T_\odot = G M_\odot/c^3 = 4.925490947\mu s$, and can be expressed in term of $f'_0$ and  $f''_0$. 
The previous equation, together with the equation of the time variation
of the orbital period in \cite{deLa_deMa2014} provides a very powerful tool to test $f(R)$-gravity with current observations from
the Parkes Pulsar Timing Array (PPTA) and, in particular, with next-generation facilities 
such as the Square-Kilometre-Array (SKA) \cite{ska1, ska2, ska3}.

 \section{Conclusion and remarks}
 \label{sei}
In this paper we have investigated the impact of the Yukawa-like gravitational potential 
on the periastron shift of an orbiting body. The Yukawa-like correction to the Newtonian potential
is a very well established result of many different alternative theories of gravity. 
Here, we have particularized our calculation to the framework of $f(R)$-gravity where
the gravitational potential assumes the functional form given in Eq.~\eqref{gravpot1}. Thus, 
the modifications due to the $f(R)$ gravity is encoded in two parameters: the strength $\delta$ and the 
scale length $\lambda$ of the Yukawa-term. 

First, we have computed the geodesic equations and we have solved 
them numerically to visually show the presence of stable orbits and the orbital precession of a test particle 
moving around a massive body. Second, we have computed an analytic formula for the periastron shift in the 
limit that the orbital radius is much lower that the scale length $\lambda$. 
Since the most suitable candidates to test the theory  are binary systems composed by a SMBH and an orbiting star,
we have computed the periastron advance particularizing the Eq.~\eqref{eq:deltaphi} for three toy models of stars orbiting
around the Galactic Center. We have illustrated our results in the Figure \ref{fig:deltaPHI_vs_delta} fixing 
$\lambda=5000 AU$. Let us remark that our results are showing the capability of the periastron shift to constrain the 
Yukawa strength once the scale length is fixed.  Then, we have generalized the expression of the periastron advance for 
a binary systems composed by two neutron stars or pulsars with comparable masses. Finally, the results showed above will represent 
a fundamental tool to be used with forthcoming observations of pulsars near the Galactic Center.

We have considered idealized systems, where the internal structure of the  
two masses and others effects that can affect their motion (like as tidal effect, dusts, etc.) have not been taken into account. 
Nevertheless, even in a realistic system, the internal structure of the stars is decoupled from the orbital motion not producing 
relevant difference in the precession. Moreover, we have particularized our plots for pulsars near to the SMBH at Galactic Center.
However, one should have in mind that finding pulsars near the SMBH is difficult  due to the relatively high density 
of free electrons in the gas around the Galactic Center.  Radio waves scatter off of these electrons, 
smearing out the sharp pulses from a pulsar in a phenomenon known as interstellar dispersion. 
Because of  searches for pulsars rely on detecting periodic bursts, 
if the pulses are smeared out over the entire pulse period, a pulsar becomes essentially undetectable.  
More stable radio pulsars in the region would allow astronomers to sample more areas of the accretion 
disk and to make accurate measurements of the curvature of space-time \cite{Eatough2013}.

Also, estimates of the pulsar population around Sgr A* range from the hundreds to the 
thousands \cite{Chennamangalam2014}.
To find these pulsars and overcome the high dispersion of pulses near the galactic center, astronomers 
will use further searches in high frequency X-rays as well as computer-intensive attempts to "de-disperse" 
observations by testing different estimates of the density of free electrons between earth and the pulsar 
at each observed point. Thus, we will soon have many more pulsars to map out the area around the SMBH. 
Forthcoming observations of the EHTC may provide a measure of the periastron shift, and other pulsar's observables  
such as the time dependence of the orbital period and the time delay, for  these pulsars. Therefore
they will provide the ultimate test for GR and alternative theories of gravity.

\section*{Acknowledgements}
We are grateful to Luciano Rezzolla for numerous helpful discussions and
comments.
M.\,D.\,L.\ is supported by the ERC Synergy Grant
``BlackHoleCam'' -- Imaging the Event Horizon of Black Holes (Grant
No.~610058). 
M.D.L. acknowledge INFN Sez. di Napoli (Iniziative Specifiche QGSKY and TEONGRAV).
I.D.M acknowledge financial supports from University of the Basque Country UPV/EHU under the program
"Convocatoria de contrataci\'{o}n para la especializaci\'{o}n de personal 
investigador doctor en la UPV/EHU 2015", from the Spanish Ministerio de
Econom\'{\i}a y Competitividad through the research project FIS2017-85076-P (MINECO/AEI/FEDER, UE),
and from  the Basque Government through the research project IT-956-16. 
This article is based upon work from COST Action CA1511 Cosmology and Astrophysics 
Network for Theoretical Advances and Training Actions (CANTATA), 
supported by COST (European Cooperation in Science and Technology).


\end{document}